\documentclass[prl,aps,twocolumn,reprint,showpacs,floatfix,superscriptaddress,longbibliography]{revtex4-1}
\usepackage[utf8]{inputenc} 
\usepackage{graphicx}
\usepackage{amsmath}

\newcommand{\be}{\begin{equation}}
\newcommand{\ee}{\end{equation}}
\newcommand{\bea}{\begin{eqnarray}}
\newcommand{\eea}{\end{eqnarray}}

\usepackage{color}

\begin{document}
\title{Tuning the dipole-dipole interaction in a quantum gas with a rotating magnetic field}

\author{Yijun Tang}
\affiliation{Department of Physics, Stanford University, Stanford CA 94305}
\affiliation{E.~L.~Ginzton Laboratory, Stanford University, Stanford CA 94305}

\author{Wil Kao}
\affiliation{E.~L.~Ginzton Laboratory, Stanford University, Stanford CA 94305}
\affiliation{Department of Applied Physics, Stanford University, Stanford CA 94305}

\author{Kuan-Yu Li}
\affiliation{E.~L.~Ginzton Laboratory, Stanford University, Stanford CA 94305}
\affiliation{Department of Applied Physics, Stanford University, Stanford CA 94305}

\author{Benjamin L. Lev}
\affiliation{Department of Physics, Stanford University, Stanford CA 94305}
\affiliation{E.~L.~Ginzton Laboratory, Stanford University, Stanford CA 94305}
\affiliation{Department of Applied Physics, Stanford University, Stanford CA 94305}

\date{\today}

\begin{abstract}
We demonstrate the tuning of the magnetic dipole-dipole interaction (DDI) within a dysprosium Bose-Einstein condensate by rapidly rotating the orientation of the atomic dipoles. The tunability of the dipolar mean-field energy manifests as a modified gas aspect ratio after time-of-flight expansion. We demonstrate that both the magnitude and the sign of the DDI can be tuned using this technique. In particular, we show that a magic rotation angle exists at which the mean-field DDI can be eliminated, and at this angle, we observe that the expansion dynamics of the condensate is close to that predicted for a non-dipolar gas. The ability to tune the strength of the DDI opens new avenues toward the creation of exotic soliton and  vortex states as well as unusual quantum lattice phases and Weyl superfluids.
\end{abstract}

 \maketitle

Recent advancements in laser cooling and trapping of highly magnetic lanthanide atoms such as dysprosium and erbium have introduced strong magnetic dipole-dipole interactions (DDI) into the toolbox of ultracold atomic physics~\cite{Dysprosium_BEC,Dysprosium_DegenerateFermiGas,Erbium_BEC,Erbium_FermiGas}. When paired with the short-ranged Van der Waals $s$-wave interaction, the long-ranged and anisotropic DDI dramatically modifies the atomic gas properties and  has enabled the exploration of a wide variety of phenomena.  These range from novel quantum liquids~\cite{Kadau:2016cb,Schmitt:2016bi,Chomaz:2016,FerrierBarbut:2016jo} and strongly correlated lattice states~\cite{dePaz:2013ff,dePaz:2016gv,Baier:2016ga}, to exotic spin dynamics~\cite{Naylor:2016bz,Lepoutre:2018vy} and the emergence of thermalization in a nearly integrable quantum gas~\cite{Tang:2017uu}. 

An even wider array of physics could be explored were one able to control the dipolar strength independent of the relative orientation of the dipoles.  For example, exotic   multidimensional bright and dark dipolar  solitons  could be observed~\cite{Pedri:2005cc,Nath:2008gt,DipolarBosonsReview} as well as exotic vortex lattices, dynamics, and interactions~\cite{Mulkerin:2013cs,Klawunn:2008ep,Cooper:2005hn}. Magnetorotons in spinor condensates~\cite{Cherng:2009bk} and the nematic susceptibility of dipolar Fermi gases~\cite{He:2008jw,Fregoso:2009cc,Zhang:2009bw,Aikawa:2014gd} could be controlled by tuning the strength of the DDI.   In optical lattices, one would be able to create tunable dipolar Luttinger liquids~\cite{Sinha:2007gx,Citro:2007gs} as well as novel quantum phases~\cite{Yi:2007dt}, including analogs of fractional quantum Hall states~\cite{Hafezi:2007gz}.  Intriguingly, Weyl superfluidity may be observable in dipolar Fermi gases by tuning the DDI~\cite{Liu:2015kq}.  Setting the DDI strength to zero has application in improving the sensitivity of atom interfermometers~\cite{Fattori:2008jj}, while tuning the strength negative may find application in the simulation of dense nuclear matter through analogies with the tensor nuclear force~\cite{Baym09}.

\begin{figure}[t!]
\centering
 \includegraphics[width=1\columnwidth]{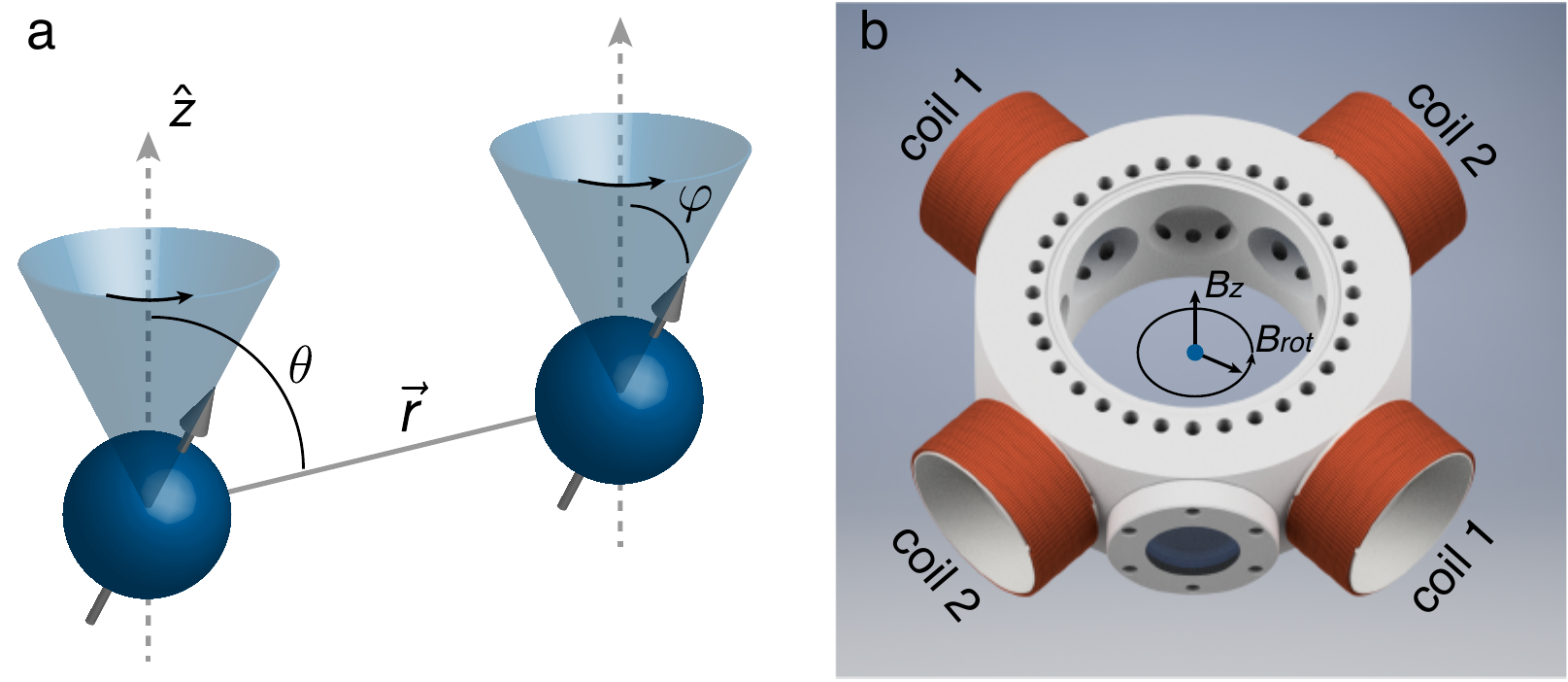}
 \caption{Tuning the DDI strength by rotating the magnetic dipoles. (a) Geometry of the rotating field technique. The dipoles are rotated along a cone centered around the $\hat{z}$ direction. (b) Schematic showing the trapping chamber and the two pairs of coils used for generating the rotating component of the magnetic field. Sizes are drawn to scale;  diameter of coils is 7~cm. The coils for generating $\hat{z}$ field and other vacuum chamber parts are not shown. The atoms are located at the center of the chamber.}
    \label{FIG:schematic}
\end{figure}

We realize a method, first proposed in 2002~\cite{Giovanazzi:2002}, to tune the DDI strength from positive to zero, and even to negative values.  Although the static DDI between two spin-polarized atoms cannot be tuned, the time-averaged DDI can be tuned by quickly rotating the dipoles.  
This provides control of the ratio  $\epsilon$ of the dipolar mean-field energy to the mean-field contact energy without the use of a Feshbach resonance to control $a_s$, the $s$-wave scattering length~\cite{Chin10_RMP}\footnote{Control of $\epsilon$ via a Feshbach resonance is often difficult in many highly magnetic atomic isotopes due to the narrowness of the resonances~\cite{Kotochigova:2014kt}.}.  This ratio is $\epsilon = \mu_0 \mu^2 m/12\pi\hbar^2 a_s$, where $\mu$ is the magnetic moment and $m$ is the mass.  The attractive component of the DDI can lead to dipolar collapse:  $\epsilon = 1 $  demarcates the boundary between mechanically stable and unstable homogeneous condensates at the mean-field level~\cite{DipolarBosonsReview}.  

Figure~\ref{FIG:schematic}(a) illustrates the geometry of the rotating dipoles.  A rotating magnetic field in the $\hat{x}$-$\hat{y}$ plane causes the dipoles to rotate at an angle $\varphi$ with respect to a static magnetic field along the $\hat{z}$-axis. Assuming cylindrically symmetric trap frequencies $\omega_x=\omega_y$ for simplicity, the time-averaged DDI between two atoms is~\cite{Giovanazzi:2002}
\begin{equation}
\langle U_\text{DDI}({\bf r},\theta,\varphi) \rangle = - \frac{\mu_0 \mu^2}{4\pi} \left( \frac{3 \cos^2{\theta} -1 }{ | {\bf r} | ^3} \right) \left( \frac{ 3 \cos^2{ \varphi } -1 }{ 2 } \right),
\label{eq:time_average_ddi}
\end{equation}
where ${\bf r}$ is the relative position vector between the two atoms, $\theta$ is the angle between ${\bf r}$ and $\hat{z}$, and $\mu=9.93\mu_\text{B}$ is the magnetic dipole moment for $^{162}$Dy, the species of atom employed for this work. This time-averaged DDI is simply the regular DDI modified by the term in the second parentheses. This term changes from 1 to $-0.5$ as $\varphi$ is tuned from $0^{\circ}$ to $90^{\circ}$ by changing the ratio of the rotation to static field strengths.  This enables the tuning of both the magnitude and the sign of the DDI. For $\varphi > \varphi_m$, even atoms sitting side-by-side experience an attractive  averaged  DDI due to the inversion of their dipoles by the rotating field.  Moreover, the DDI vanishes for any $\theta$, i.e., any pair of atoms in the gas, at the so-called magic angle $\varphi_m=54.7^{\circ}$.  We note that an alternative method for reducing the strength of the DDI---spin-polarizing in $|m_F|<F$ Zeeman substates---unfortunately leads to gas heating and/or atom loss from  dipolar relaxation~\cite{Hensler:2003,Pasquiou:2010,Burdick:2015bx}.

In this work, we prepare $^{162}$Dy BECs with $2.0(2)\times10^4$ atoms in the absolute ground Zeeman sublevel $m_J=-8$ ($J=8$).
The BECs are created by evaporative cooling in crossed 1064-nm optical dipole traps (ODT).  The procedure is similar to that described in a previous publication~\cite{Tang2015}.  The present experiment differs only in that instead of loading atoms from the magneto-optical trap using a spatially dithered circular ODT beam, we now use a stationary elliptical ODT with a horizontal waist of 73(3)~$\mu$m and a vertical waist of 19(2)~$\mu$m. 

The rapid rotation of the atomic dipoles is realized by rotating a bias magnetic field at $\omega_r=2\pi\times 1$~kHz.  This is chosen to be fast compared to the trap frequencies [$\omega_x$,$\omega_y$,$\omega_z$] = $2\pi\times$[73(1),37(2),74(1)]~Hz to avoid parametric heating, but is slow compared to the Larmor frequency 1.55~MHz to ensure that the rotation is adiabatic. The rotating field consists of a static component along $\hat{z}$ and a rotating component in the $x$-$y$ plane generated by a pair of coils driven 90$^{\circ}$ out-of-phase using two bipolar current sources, as illustrated in Fig.~\ref{FIG:schematic}(b). The total field as a function of time $t$ can be written as ${\bf B}(t) = B_\text{rot} \left[ \cos{(\omega_r t+\pi/4)} \hat{x} + \sin{(\omega_r t+\pi/4)} \hat{y} \right] + B_z \hat{z}$, where the total magnitude $B=\sqrt{ B_\text{rot}^2 + B_z^2 }$ is fixed at 0.89(2)~G~\footnote{The error in this number is quoted as the worst case error.   This worst case  occurs at $\varphi=90^\circ$, since the uncertainty in  $B_\text{rot}$ is larger than in $B_z$.}, away from any Feshbach resonances~\cite{Baumann:2014ey}, and the rotation angle is related to the magnitude of the two components by $\tan{\varphi}=B_\text{rot}/B_z$. The vertical field $B_z$ is provided by a pair of coils in the $\hat{z}$ direction and is not shown in Fig.~\ref{FIG:schematic}(b). The angle $\varphi$ is  controlled using a calibration procedure that corrects for the effect of eddy currents.  We now describe the calibration.

\begin{figure}[t!]
\centering
 \includegraphics[width=1\columnwidth]{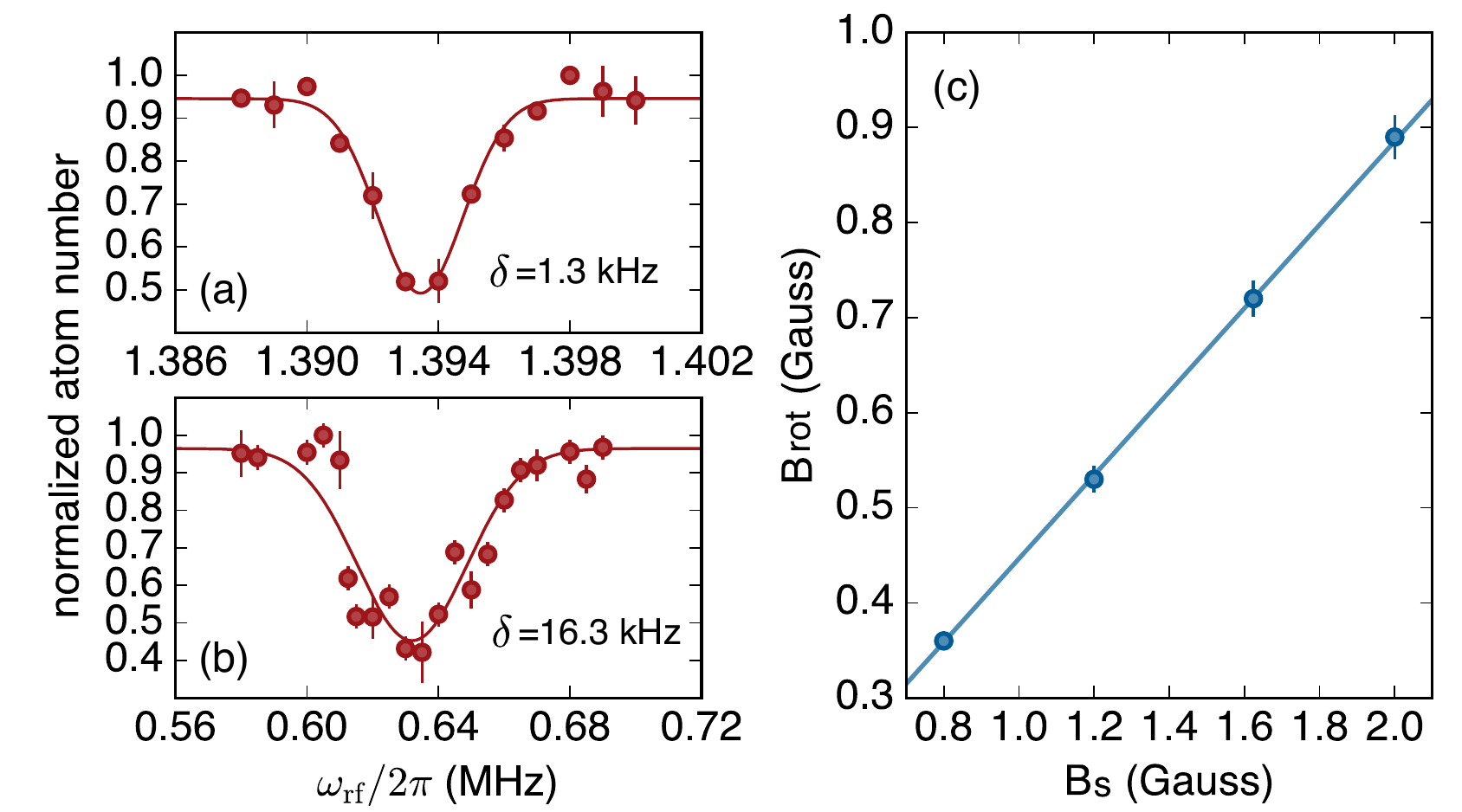}
 \caption{(a) Atom-loss spectrum from a rf-spectroscopy measurement at a DC field $B_s=0.802$~G. (b) Atom-loss spectrum from a rf-spectroscopy measurement for a rotating field $B_\text{rot}$ generated with AC current of the same amplitude. The resonance shifts to a lower frequency due to effects of eddy currents. The resonance width increases from 1.3~kHz (0.7~mG) to 16.3~kHz (9.4~mG) due to residual fluctuations of the rotating field amplitude. (c) Measured linear dependence between $B_\text{rot}$ and $B_s$. Error bars represent one standard error.}
    \label{FIG:calibration}
\end{figure}

Because the coils generating the rotating component of the field are mounted outside the stainless steel vacuum chamber, the magnitude of the rotating component $B_\text{rot}$ is reduced due to eddy currents compared to a static field $B_s$ generated by driving the coils with the equivalent DC current. We calibrate the effect of eddy currents by measuring $B_\text{rot}$ at different $B_s$. The field magnitude is measured using rf-spectroscopy, where we drive the atoms with a single-tone rf-field at frequency $\omega_\text{rf}$. When $\omega_\text{rf}$ matches the Zeeman splitting, the atoms are transferred to higher Zeeman states and subsequently dipolar relax.  This  causes rapid atom loss, which heralds the resonance~\cite{Hensler:2003,Pasquiou:2010,Burdick:2015bx}.  The Zeeman splitting is 1.7378~MHz/G for  bosonic dysprosium~\cite{Martin:1978}. The atom loss spectra of a typical set of rotating $B_\text{rot}$ and static $B_s$ fields  are shown in Fig.~\ref{FIG:calibration}. The magnitude of the field can be determined from the central location of the atom-loss resonance, and the stability of the field can be determined from the resonance linewidth. Figure~\ref{FIG:calibration}(a) shows the spectrum for a static field. The resonance center is located at $\omega_\text{rf}=1.393$~MHz, corresponding to 0.802~G, and the linewidth, defined as the standard deviation of a Gaussian fit, is 1.3~kHz, equivalent to 0.7~mG. When the coils are driven with AC current of the same amplitude, the resulting rf-spectrum is shown in Fig.~\ref{FIG:calibration}(b). The magnitude of this rotating field is reduced to $B_\text{rot}=0.364$~G by eddy currents and the linewidth is broadened to 9.4~mG. This broadening provides a measure of the field's amplitude fluctuations while the field rotates. The magnitude of the fluctuation in this case is 2.6\%. We measured a total of four sets of $B_s$ and $B_\text{rot}$, and the results are shown in Fig.~\ref{FIG:calibration}(c). We observe a linear dependence within this field range: $B_\text{rot}=\alpha B_s$, where $\alpha=0.445(6)$. By using this calibration, we are able to determine the amplitude of the AC current required to produce a given rotation angle $\varphi$.

To study the manifestation of the time-averaged DDI, we measure the change in the BEC mean-field energy due to the rotating field by observing the change in aspect ratio (AR) of the BEC. We first prepare a BEC in a static bias field along $\hat{z}$. We then ramp the currents in the $\hat{z}$-coil and coil 1 to rotate the field from $\hat{z}$ to the ${\bf B}(0)$ configuration, setting the initial condition for the rotating field. After 10 cycles of rotation, we suddenly (in $<$200~$\mu$s) turn off the ODTs and let the BEC expand.  We continue to rotate the fields for the first 5~ms of the time-of-flight (TOF) expansion; afterwards, the density of the atomic gas is low enough that the interactions no longer affect expansion dynamics and we can safely turn off the rotating fields without affecting the gas AR.  During this first 5-ms of TOF, the gas falls 125~$\mu$m under gravity. At this displacement, the gas experiences a transverse field generated by coils 1 and 2 that is only 0.1\% of the axial field: the variation of the rotation angle $\varphi$ is negligible during the initial 5~ms of TOF. We then perform absorption imaging on the resonant 421-nm transition along the $\hat{y}$-direction to measure the momentum distribution in the $x$-$z$ plane. We fit  1D integrated density profiles  along both $\hat{x}$ and $\hat{z}$ to  integrated Thomas-Fermi distributions $n(r_i)\sim [\max(1-r_i^2/R_i^2,0)]^2$. The AR is defined as the ratio of the extracted Thomas-Fermi radii $R_z/R_x$.

\begin{figure}[t!]
\centering
 \includegraphics[width=1\columnwidth]{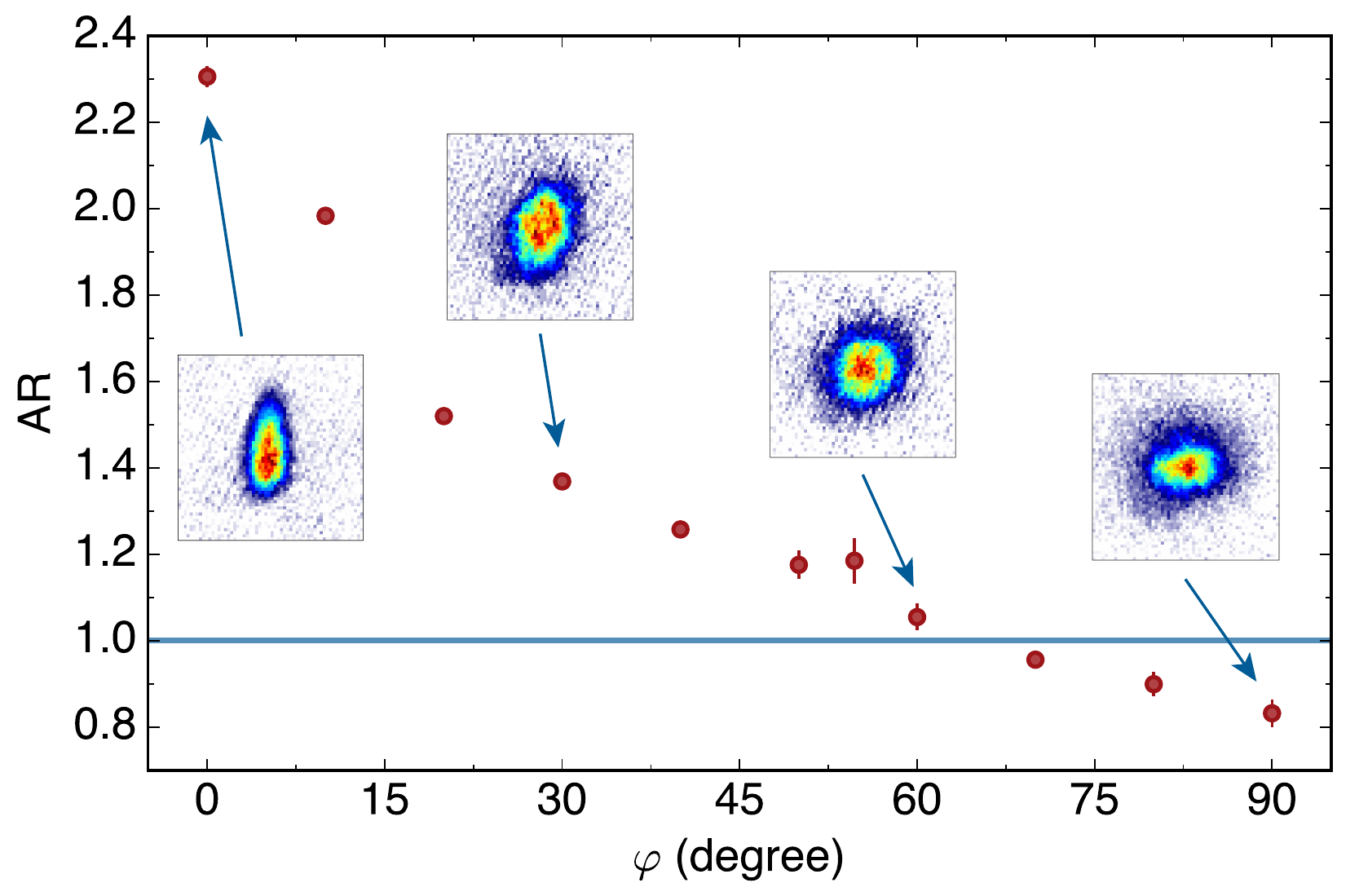}
 \caption{Aspect ratio (AR) of the BEC after 19~ms of TOF expansion as a function of rotation angle $\varphi$. Sample single-shot absorption images for $\varphi=0^\circ,30^\circ,60^\circ,90^\circ$ are shown in insets. The AR can be tuned from $\sim$2.3 in a static $\hat{z}$ field to below unity in a fully rotating field at $\varphi=90^\circ$. The $\varphi=0^\circ$ case corresponds to a static $1.580(5)$~G $\hat{z}$ field. Error bars are standard error from three measurements.}
    \label{FIG:ar_vs_angle}
\end{figure}

The AR of the BEC after 19~ms of TOF is shown in Fig.~\ref{FIG:ar_vs_angle} for different $\varphi$. We observe that the AR monotonically decreases from $\sim$2.3 at $\varphi=0^\circ$, corresponding to a static $\hat{z}$ field where the DDI is maximally repulsive, to below unity at $\varphi=90^\circ$. Sample single-shot absorption images for $\varphi=0^\circ,30^\circ,60^\circ,90^\circ$ are shown in insets of Fig.~\ref{FIG:ar_vs_angle}. We note that the Thomas-Fermi radius of a non-dipolar BEC evolves in a free expansion according to $R_i(t)=\lambda_i(t)R_i(0)$, where $R_i(0)$ is the in-trap Thomas-Fermi radius and the scaling factor $\lambda_i$ can be found by solving $\ddot{\lambda_i}=\omega_i^2/(\lambda_i\lambda_x\lambda_y\lambda_z)$ with initial condition $\lambda_i(0)=1$, where $i=x,y,z$~\cite{Castin:1996}. For the trap employed in this work, we have $\omega_x\approx\omega_z$ and therefore the BEC AR should simply be equal to one in the absence of the DDI.  However, the fact that  AR does not equal one in our experiment is due to the DDI~\cite{DipolarBosonsReview}.   The observed reduction of  AR with rotation angle---even to below unity---is evidence that the DDI can be tuned, as expected from Eq.~(\ref{eq:time_average_ddi}). We note that the AR scaling with $\varphi$ is not exactly what Eq.~\ref{eq:time_average_ddi} predicts. For example, the AR at $\varphi$ does not equal -0.5$\times$ that at $\varphi=0$.  We find that  mean-field treatments of the dipolar BEC expansion do not adequately fit our data with or without rotation. Further work must be  done to extend such treatments to account for beyond mean-field effects and/or hydrodynamic effects in the early expansion~\cite{Kadau:2016cb,Tang:2016prl}.

\begin{figure}[t!]
\centering
 \includegraphics[width=1\columnwidth]{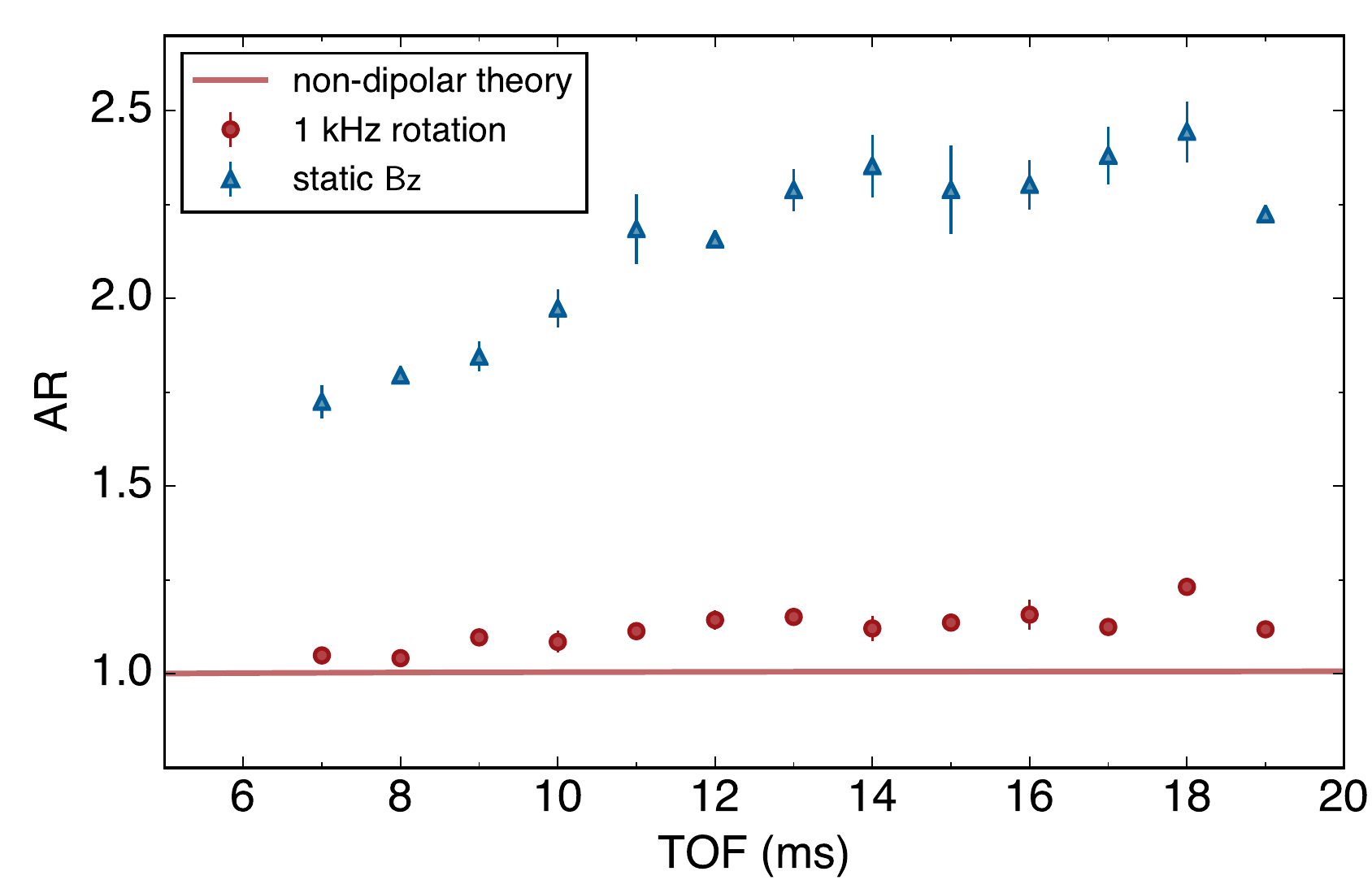}
 \caption{AR of the BEC as a function of TOF. Triangle: expansion in a 1.580(5)-G static $\hat{z}$ field (i.e., $\varphi = 0^\circ$). Circle: expansion in a 0.89(2) field rotating at the magic angle $\varphi_m=54.7^\circ$. Solid line: non-dipolar theory prediction, which is equal to unity at all times for cylindrically symmetric trap parameters.  The employed trap is approximately cylindrically symmetric.  Error bars are standard error from three measurements.}
    \label{FIG:ar_vs_tof}
\end{figure}

We also compared the evolution of AR as a function of TOF for BECs  in a static field and in fields rotating at the magic angle $\varphi_m$ (at which the time-averaged DDI should be zero). The results are shown in Fig.~\ref{FIG:ar_vs_tof} for TOFs spanning 7~ms to 19~ms at 1-ms intervals. The BEC gas is too dense for reliable absorption imaging earlier than 7~ms of TOF. As expected for a dipolar gas in a symmetric trap, we observe that  the BEC  is highly anisotropic at 7~ms of TOF in a static $\hat{z}$ field (i.e., $\varphi = 0^\circ$). The AR asymptotes to $\sim$2.3. However, the AR remains near unity when expanding in a field rotating at the magic angle.   This concurs with expectations for a non-dipolar BEC, suggesting that the rotating field succeeds in nearly eliminating the dipolar mean-field energy at the magic angle. Equation~(\ref{eq:time_average_ddi}) is derived under the assumption of cylindrical symmetry about $\hat{z}$: The residual deviation from unity AR may be due to the lack of this  cylindrical symmetry  in the trap employed. 

Above rotation rates of a few hundred Hz, i.e., well above the trap frequencies, we observed that the $1/e$ population lifetime of our BEC reached a maximum of $\sim$160~ms~\footnote{We were unable to measure lifetimes at rotation rates  faster than $\sim$1~kHz due to power supply limitations and increasingly large eddy currents.}.  While this lifetime is  sufficiently long for many experiments, it is  one or two orders of magnitude shorter than a Dy BEC in static fields in our apparatus.  The atom loss and heating are likely due to  residual field gradients that lead to a parametric motional excitation associated with the rotating component $B_\text{rot}$. As shown in Fig.~\ref{FIG:schematic}(b), drawn to scale, the two coils are not in strict Helmholtz coil configuration, leading to non-negligible field gradients. Eddy currents in the vacuum parts could also lead to heating, but this effect cannot be controlled or separately measured in our present apparatus. We expect that by placing two pairs of orthogonal Helmholtz coils inside vacuum, or outside a glass cell,  one could significantly improve the lifetime of the BEC in a rotating field. 

In summary, we realized a scheme to tune the averaged DDI strength in a dipolar BEC. This was accomplished by rapidly rotating a magnetic field. We demonstrate that the AR of the BEC after long TOF can be tuned from 2.3 to below unity, confirming the expectation from Eq.~(\ref{eq:time_average_ddi}), introduced in Ref.~\cite{Giovanazzi:2002}, that both the magnitude and sign of the DDI can be tuned by rotating the dipoles at different angles $\varphi$. Furthermore, at the magic rotation angle $\varphi_m=54.7^\circ$, expansion dynamics of our dysprosium BEC is similar to that of a non-dipolar gas, demonstrating that the DDI can be nearly turned off in rotating fields.  This work shows that a new tool---the tuning of the DDI, and consequently, $\epsilon$---is readily available to control atomic interactions for the propose of creating exotic quantum many-body systems.

We acknowledge support from AFOSR (FA9550-17-1-0266) and NSF (PHY-1707336). WK acknowledges support from NSERC Postgraduate Scholarship-Doctoral.


\begin{thebibliography}{46}%
\makeatletter
\providecommand \@ifxundefined [1]{%
 \@ifx{#1\undefined}
}%
\providecommand \@ifnum [1]{%
 \ifnum #1\expandafter \@firstoftwo
 \else \expandafter \@secondoftwo
 \fi
}%
\providecommand \@ifx [1]{%
 \ifx #1\expandafter \@firstoftwo
 \else \expandafter \@secondoftwo
 \fi
}%
\providecommand \natexlab [1]{#1}%
\providecommand \enquote  [1]{``#1''}%
\providecommand \bibnamefont  [1]{#1}%
\providecommand \bibfnamefont [1]{#1}%
\providecommand \citenamefont [1]{#1}%
\providecommand \href@noop [0]{\@secondoftwo}%
\providecommand \href [0]{\begingroup \@sanitize@url \@href}%
\providecommand \@href[1]{\@@startlink{#1}\@@href}%
\providecommand \@@href[1]{\endgroup#1\@@endlink}%
\providecommand \@sanitize@url [0]{\catcode `\\12\catcode `\$12\catcode
  `\&12\catcode `\#12\catcode `\^12\catcode `\_12\catcode `\%12\relax}%
\providecommand \@@startlink[1]{}%
\providecommand \@@endlink[0]{}%
\providecommand \url  [0]{\begingroup\@sanitize@url \@url }%
\providecommand \@url [1]{\endgroup\@href {#1}{\urlprefix }}%
\providecommand \urlprefix  [0]{URL }%
\providecommand \Eprint [0]{\href }%
\providecommand \doibase [0]{http://dx.doi.org/}%
\providecommand \selectlanguage [0]{\@gobble}%
\providecommand \bibinfo  [0]{\@secondoftwo}%
\providecommand \bibfield  [0]{\@secondoftwo}%
\providecommand \translation [1]{[#1]}%
\providecommand \BibitemOpen [0]{}%
\providecommand \bibitemStop [0]{}%
\providecommand \bibitemNoStop [0]{.\EOS\space}%
\providecommand \EOS [0]{\spacefactor3000\relax}%
\providecommand \BibitemShut  [1]{\csname bibitem#1\endcsname}%
\let\auto@bib@innerbib\@empty
\bibitem [{\citenamefont {Lu}\ \emph {et~al.}(2011)\citenamefont {Lu},
  \citenamefont {Burdick}, \citenamefont {Youn},\ and\ \citenamefont
  {Lev}}]{Dysprosium_BEC}%
  \BibitemOpen
  \bibfield  {author} {\bibinfo {author} {\bibfnamefont {M.}~\bibnamefont
  {Lu}}, \bibinfo {author} {\bibfnamefont {N.~Q.}\ \bibnamefont {Burdick}},
  \bibinfo {author} {\bibfnamefont {S.~H.}\ \bibnamefont {Youn}}, \ and\
  \bibinfo {author} {\bibfnamefont {B.~L.}\ \bibnamefont {Lev}},\ }\bibfield
  {title} {\enquote {\bibinfo {title} {Strongly dipolar {Bose-Einstein}
  condensate of dysprosium},}\ }\href {\doibase 10.1103/PhysRevLett.107.190401}
  {\bibfield  {journal} {\bibinfo  {journal} {Phys. Rev. Lett.}\ }\textbf
  {\bibinfo {volume} {107}},\ \bibinfo {pages} {190401} (\bibinfo {year}
  {2011})}\BibitemShut {NoStop}%
\bibitem [{\citenamefont {Lu}\ \emph {et~al.}(2012)\citenamefont {Lu},
  \citenamefont {Burdick},\ and\ \citenamefont
  {Lev}}]{Dysprosium_DegenerateFermiGas}%
  \BibitemOpen
  \bibfield  {author} {\bibinfo {author} {\bibfnamefont {M.}~\bibnamefont
  {Lu}}, \bibinfo {author} {\bibfnamefont {N.~Q.}\ \bibnamefont {Burdick}}, \
  and\ \bibinfo {author} {\bibfnamefont {B.~L.}\ \bibnamefont {Lev}},\
  }\bibfield  {title} {\enquote {\bibinfo {title} {Quantum degenerate dipolar
  fermi gas},}\ }\href {\doibase 10.1103/PhysRevLett.108.215301} {\bibfield
  {journal} {\bibinfo  {journal} {Phys. Rev. Lett.}\ }\textbf {\bibinfo
  {volume} {108}},\ \bibinfo {pages} {215301} (\bibinfo {year}
  {2012})}\BibitemShut {NoStop}%
\bibitem [{\citenamefont {Aikawa}\ \emph {et~al.}(2012)\citenamefont {Aikawa},
  \citenamefont {Frisch}, \citenamefont {Mark}, \citenamefont {Baier},
  \citenamefont {Rietzler}, \citenamefont {Grimm},\ and\ \citenamefont
  {Ferlaino}}]{Erbium_BEC}%
  \BibitemOpen
  \bibfield  {author} {\bibinfo {author} {\bibfnamefont {K.}~\bibnamefont
  {Aikawa}}, \bibinfo {author} {\bibfnamefont {A.}~\bibnamefont {Frisch}},
  \bibinfo {author} {\bibfnamefont {M.}~\bibnamefont {Mark}}, \bibinfo {author}
  {\bibfnamefont {S.}~\bibnamefont {Baier}}, \bibinfo {author} {\bibfnamefont
  {A.}~\bibnamefont {Rietzler}}, \bibinfo {author} {\bibfnamefont
  {R.}~\bibnamefont {Grimm}}, \ and\ \bibinfo {author} {\bibfnamefont
  {F.}~\bibnamefont {Ferlaino}},\ }\bibfield  {title} {\enquote {\bibinfo
  {title} {{Bose-Einstein} condensation of erbium},}\ }\href {\doibase
  10.1103/PhysRevLett.108.210401} {\bibfield  {journal} {\bibinfo  {journal}
  {Phys. Rev. Lett.}\ }\textbf {\bibinfo {volume} {108}},\ \bibinfo {pages}
  {210401} (\bibinfo {year} {2012})}\BibitemShut {NoStop}%
\bibitem [{\citenamefont {Aikawa}\ \emph
  {et~al.}(2014{\natexlab{a}})\citenamefont {Aikawa}, \citenamefont {Frisch},
  \citenamefont {Mark}, \citenamefont {Baier}, \citenamefont {Grimm},\ and\
  \citenamefont {Ferlaino}}]{Erbium_FermiGas}%
  \BibitemOpen
  \bibfield  {author} {\bibinfo {author} {\bibfnamefont {K.}~\bibnamefont
  {Aikawa}}, \bibinfo {author} {\bibfnamefont {A.}~\bibnamefont {Frisch}},
  \bibinfo {author} {\bibfnamefont {M.}~\bibnamefont {Mark}}, \bibinfo {author}
  {\bibfnamefont {S.}~\bibnamefont {Baier}}, \bibinfo {author} {\bibfnamefont
  {R.}~\bibnamefont {Grimm}}, \ and\ \bibinfo {author} {\bibfnamefont
  {F.}~\bibnamefont {Ferlaino}},\ }\bibfield  {title} {\enquote {\bibinfo
  {title} {Reaching fermi degeneracy via universal dipolar scattering},}\
  }\href {\doibase 10.1103/PhysRevLett.112.010404} {\bibfield  {journal}
  {\bibinfo  {journal} {Phys. Rev. Lett.}\ }\textbf {\bibinfo {volume} {112}},\
  \bibinfo {pages} {010404} (\bibinfo {year} {2014}{\natexlab{a}})}\BibitemShut
  {NoStop}%
\bibitem [{\citenamefont {Kadau}\ \emph {et~al.}(2016)\citenamefont {Kadau},
  \citenamefont {Schmitt}, \citenamefont {Wenzel}, \citenamefont {Wink},
  \citenamefont {Maier}, \citenamefont {Ferrier-Barbut},\ and\ \citenamefont
  {Pfau}}]{Kadau:2016cb}%
  \BibitemOpen
  \bibfield  {author} {\bibinfo {author} {\bibfnamefont {H.}~\bibnamefont
  {Kadau}}, \bibinfo {author} {\bibfnamefont {M.}~\bibnamefont {Schmitt}},
  \bibinfo {author} {\bibfnamefont {M.}~\bibnamefont {Wenzel}}, \bibinfo
  {author} {\bibfnamefont {C.}~\bibnamefont {Wink}}, \bibinfo {author}
  {\bibfnamefont {T.}~\bibnamefont {Maier}}, \bibinfo {author} {\bibfnamefont
  {I.}~\bibnamefont {Ferrier-Barbut}}, \ and\ \bibinfo {author} {\bibfnamefont
  {T.}~\bibnamefont {Pfau}},\ }\bibfield  {title} {\enquote {\bibinfo {title}
  {{Observing the Rosensweig instability of a quantum ferrofluid}},}\
  }\href@noop {} {\bibfield  {journal} {\bibinfo  {journal} {Nature}\ }\textbf
  {\bibinfo {volume} {530}},\ \bibinfo {pages} {194--197} (\bibinfo {year}
  {2016})}\BibitemShut {NoStop}%
\bibitem [{\citenamefont {Schmitt}\ \emph {et~al.}(2016)\citenamefont
  {Schmitt}, \citenamefont {Wenzel}, \citenamefont {B{\"o}ttcher},
  \citenamefont {Ferrier-Barbut},\ and\ \citenamefont {Pfau}}]{Schmitt:2016bi}%
  \BibitemOpen
  \bibfield  {author} {\bibinfo {author} {\bibfnamefont {M.}~\bibnamefont
  {Schmitt}}, \bibinfo {author} {\bibfnamefont {M.}~\bibnamefont {Wenzel}},
  \bibinfo {author} {\bibfnamefont {F.}~\bibnamefont {B{\"o}ttcher}}, \bibinfo
  {author} {\bibfnamefont {I.}~\bibnamefont {Ferrier-Barbut}}, \ and\ \bibinfo
  {author} {\bibfnamefont {T.}~\bibnamefont {Pfau}},\ }\bibfield  {title}
  {\enquote {\bibinfo {title} {{Self-bound droplets of a dilute magnetic
  quantum liquid}},}\ }\href@noop {} {\bibfield  {journal} {\bibinfo  {journal}
  {Nature}\ }\textbf {\bibinfo {volume} {539}},\ \bibinfo {pages} {259--262}
  (\bibinfo {year} {2016})}\BibitemShut {NoStop}%
\bibitem [{\citenamefont {Chomaz}\ \emph {et~al.}(2016)\citenamefont {Chomaz},
  \citenamefont {Baier}, \citenamefont {Petter}, \citenamefont {Mark},
  \citenamefont {W\"achtler}, \citenamefont {Santos},\ and\ \citenamefont
  {Ferlaino}}]{Chomaz:2016}%
  \BibitemOpen
  \bibfield  {author} {\bibinfo {author} {\bibfnamefont {L.}~\bibnamefont
  {Chomaz}}, \bibinfo {author} {\bibfnamefont {S.}~\bibnamefont {Baier}},
  \bibinfo {author} {\bibfnamefont {D.}~\bibnamefont {Petter}}, \bibinfo
  {author} {\bibfnamefont {M.~J.}\ \bibnamefont {Mark}}, \bibinfo {author}
  {\bibfnamefont {F.}~\bibnamefont {W\"achtler}}, \bibinfo {author}
  {\bibfnamefont {L.}~\bibnamefont {Santos}}, \ and\ \bibinfo {author}
  {\bibfnamefont {F.}~\bibnamefont {Ferlaino}},\ }\bibfield  {title} {\enquote
  {\bibinfo {title} {Quantum-fluctuation-driven crossover from a dilute
  {Bose-Einstein} condensate to a macrodroplet in a dipolar quantum fluid},}\
  }\href {\doibase 10.1103/PhysRevX.6.041039} {\bibfield  {journal} {\bibinfo
  {journal} {Phys. Rev. X}\ }\textbf {\bibinfo {volume} {6}},\ \bibinfo {pages}
  {041039} (\bibinfo {year} {2016})}\BibitemShut {NoStop}%
\bibitem [{\citenamefont {Ferrier-Barbut}\ \emph {et~al.}(2016)\citenamefont
  {Ferrier-Barbut}, \citenamefont {Kadau}, \citenamefont {Schmitt},
  \citenamefont {Wenzel},\ and\ \citenamefont {Pfau}}]{FerrierBarbut:2016jo}%
  \BibitemOpen
  \bibfield  {author} {\bibinfo {author} {\bibfnamefont {I.}~\bibnamefont
  {Ferrier-Barbut}}, \bibinfo {author} {\bibfnamefont {H.}~\bibnamefont
  {Kadau}}, \bibinfo {author} {\bibfnamefont {M.}~\bibnamefont {Schmitt}},
  \bibinfo {author} {\bibfnamefont {M.}~\bibnamefont {Wenzel}}, \ and\ \bibinfo
  {author} {\bibfnamefont {T.}~\bibnamefont {Pfau}},\ }\bibfield  {title}
  {\enquote {\bibinfo {title} {Observation of quantum droplets in a strongly
  dipolar bose gas},}\ }\href@noop {} {\bibfield  {journal} {\bibinfo
  {journal} {Phys. Rev. Lett.}\ }\textbf {\bibinfo {volume} {116}},\ \bibinfo
  {pages} {215301} (\bibinfo {year} {2016})}\BibitemShut {NoStop}%
\bibitem [{\citenamefont {de~Paz}\ \emph {et~al.}(2013)\citenamefont {de~Paz},
  \citenamefont {Sharma}, \citenamefont {Chotia}, \citenamefont {Mar{\'e}chal},
  \citenamefont {Huckans}, \citenamefont {Pedri}, \citenamefont {Santos},
  \citenamefont {Gorceix}, \citenamefont {Vernac},\ and\ \citenamefont
  {Laburthe-Tolra}}]{dePaz:2013ff}%
  \BibitemOpen
  \bibfield  {author} {\bibinfo {author} {\bibfnamefont {A.}~\bibnamefont
  {de~Paz}}, \bibinfo {author} {\bibfnamefont {A.}~\bibnamefont {Sharma}},
  \bibinfo {author} {\bibfnamefont {A.}~\bibnamefont {Chotia}}, \bibinfo
  {author} {\bibfnamefont {E.}~\bibnamefont {Mar{\'e}chal}}, \bibinfo {author}
  {\bibfnamefont {J.~H.}\ \bibnamefont {Huckans}}, \bibinfo {author}
  {\bibfnamefont {P.}~\bibnamefont {Pedri}}, \bibinfo {author} {\bibfnamefont
  {L.}~\bibnamefont {Santos}}, \bibinfo {author} {\bibfnamefont
  {O.}~\bibnamefont {Gorceix}}, \bibinfo {author} {\bibfnamefont
  {L.}~\bibnamefont {Vernac}}, \ and\ \bibinfo {author} {\bibfnamefont
  {B.}~\bibnamefont {Laburthe-Tolra}},\ }\bibfield  {title} {\enquote {\bibinfo
  {title} {{Nonequilibrium Quantum Magnetism in a Dipolar Lattice Gas}},}\
  }\href@noop {} {\bibfield  {journal} {\bibinfo  {journal} {Phys. Rev. Lett.}\
  }\textbf {\bibinfo {volume} {111}},\ \bibinfo {pages} {185305} (\bibinfo
  {year} {2013})}\BibitemShut {NoStop}%
\bibitem [{\citenamefont {de~Paz}\ \emph {et~al.}(2016)\citenamefont {de~Paz},
  \citenamefont {Pedri}, \citenamefont {Sharma}, \citenamefont {Efremov},
  \citenamefont {Naylor}, \citenamefont {Gorceix}, \citenamefont
  {Mar{\'e}chal}, \citenamefont {Vernac},\ and\ \citenamefont
  {Laburthe-Tolra}}]{dePaz:2016gv}%
  \BibitemOpen
  \bibfield  {author} {\bibinfo {author} {\bibfnamefont {A.}~\bibnamefont
  {de~Paz}}, \bibinfo {author} {\bibfnamefont {P.}~\bibnamefont {Pedri}},
  \bibinfo {author} {\bibfnamefont {A.}~\bibnamefont {Sharma}}, \bibinfo
  {author} {\bibfnamefont {M.}~\bibnamefont {Efremov}}, \bibinfo {author}
  {\bibfnamefont {B.}~\bibnamefont {Naylor}}, \bibinfo {author} {\bibfnamefont
  {O.}~\bibnamefont {Gorceix}}, \bibinfo {author} {\bibfnamefont
  {E.}~\bibnamefont {Mar{\'e}chal}}, \bibinfo {author} {\bibfnamefont
  {L.}~\bibnamefont {Vernac}}, \ and\ \bibinfo {author} {\bibfnamefont
  {B.}~\bibnamefont {Laburthe-Tolra}},\ }\bibfield  {title} {\enquote {\bibinfo
  {title} {{Probing spin dynamics from the Mott insulating to the superfluid
  regime in a dipolar lattice gas}},}\ }\href@noop {} {\bibfield  {journal}
  {\bibinfo  {journal} {Phys. Rev. A}\ }\textbf {\bibinfo {volume} {93}},\
  \bibinfo {pages} {021603} (\bibinfo {year} {2016})}\BibitemShut {NoStop}%
\bibitem [{\citenamefont {Baier}\ \emph {et~al.}(2016)\citenamefont {Baier},
  \citenamefont {Mark}, \citenamefont {Petter}, \citenamefont {Aikawa},
  \citenamefont {Chomaz}, \citenamefont {Cai}, \citenamefont {Baranov},
  \citenamefont {Zoller},\ and\ \citenamefont {Ferlaino}}]{Baier:2016ga}%
  \BibitemOpen
  \bibfield  {author} {\bibinfo {author} {\bibfnamefont {S.}~\bibnamefont
  {Baier}}, \bibinfo {author} {\bibfnamefont {M.~J.}\ \bibnamefont {Mark}},
  \bibinfo {author} {\bibfnamefont {D.}~\bibnamefont {Petter}}, \bibinfo
  {author} {\bibfnamefont {K.}~\bibnamefont {Aikawa}}, \bibinfo {author}
  {\bibfnamefont {L.}~\bibnamefont {Chomaz}}, \bibinfo {author} {\bibfnamefont
  {Z.}~\bibnamefont {Cai}}, \bibinfo {author} {\bibfnamefont {M.}~\bibnamefont
  {Baranov}}, \bibinfo {author} {\bibfnamefont {P.}~\bibnamefont {Zoller}}, \
  and\ \bibinfo {author} {\bibfnamefont {F.}~\bibnamefont {Ferlaino}},\
  }\bibfield  {title} {\enquote {\bibinfo {title} {{Extended Bose-Hubbard
  models with ultracold magnetic atoms}},}\ }\href@noop {} {\bibfield
  {journal} {\bibinfo  {journal} {Science}\ }\textbf {\bibinfo {volume}
  {352}},\ \bibinfo {pages} {201} (\bibinfo {year} {2016})}\BibitemShut
  {NoStop}%
\bibitem [{\citenamefont {Naylor}\ \emph {et~al.}(2016)\citenamefont {Naylor},
  \citenamefont {Brewczyk}, \citenamefont {Gajda}, \citenamefont {Gorceix},
  \citenamefont {Mar{\'e}chal}, \citenamefont {Vernac},\ and\ \citenamefont
  {Laburthe-Tolra}}]{Naylor:2016bz}%
  \BibitemOpen
  \bibfield  {author} {\bibinfo {author} {\bibfnamefont {B.}~\bibnamefont
  {Naylor}}, \bibinfo {author} {\bibfnamefont {M.}~\bibnamefont {Brewczyk}},
  \bibinfo {author} {\bibfnamefont {M.}~\bibnamefont {Gajda}}, \bibinfo
  {author} {\bibfnamefont {O.}~\bibnamefont {Gorceix}}, \bibinfo {author}
  {\bibfnamefont {E.}~\bibnamefont {Mar{\'e}chal}}, \bibinfo {author}
  {\bibfnamefont {L.}~\bibnamefont {Vernac}}, \ and\ \bibinfo {author}
  {\bibfnamefont {B.}~\bibnamefont {Laburthe-Tolra}},\ }\bibfield  {title}
  {\enquote {\bibinfo {title} {Competition between {Bose-Einstein} condensation
  and spin dynamics},}\ }\href@noop {} {\bibfield  {journal} {\bibinfo
  {journal} {Phys. Rev. Lett.}\ }\textbf {\bibinfo {volume} {117}},\ \bibinfo
  {pages} {185302} (\bibinfo {year} {2016})}\BibitemShut {NoStop}%
\bibitem [{\citenamefont {Lepoutre}\ \emph {et~al.}(2018)\citenamefont
  {Lepoutre}, \citenamefont {Schachenmayer}, \citenamefont {Gabardos},
  \citenamefont {Zhu}, \citenamefont {Naylor}, \citenamefont {Mar{\'e}chal},
  \citenamefont {Gorceix}, \citenamefont {Rey}, \citenamefont {Vernac},\ and\
  \citenamefont {Laburthe-Tolra}}]{Lepoutre:2018vy}%
  \BibitemOpen
  \bibfield  {author} {\bibinfo {author} {\bibfnamefont {S.}~\bibnamefont
  {Lepoutre}}, \bibinfo {author} {\bibfnamefont {J.}~\bibnamefont
  {Schachenmayer}}, \bibinfo {author} {\bibfnamefont {L.}~\bibnamefont
  {Gabardos}}, \bibinfo {author} {\bibfnamefont {B.}~\bibnamefont {Zhu}},
  \bibinfo {author} {\bibfnamefont {B.}~\bibnamefont {Naylor}}, \bibinfo
  {author} {\bibfnamefont {E.}~\bibnamefont {Mar{\'e}chal}}, \bibinfo {author}
  {\bibfnamefont {O.}~\bibnamefont {Gorceix}}, \bibinfo {author} {\bibfnamefont
  {A.~M.}\ \bibnamefont {Rey}}, \bibinfo {author} {\bibfnamefont
  {L.}~\bibnamefont {Vernac}}, \ and\ \bibinfo {author} {\bibfnamefont
  {B.}~\bibnamefont {Laburthe-Tolra}},\ }\bibfield  {title} {\enquote {\bibinfo
  {title} {{Exploring out-of-equilibrium quantum magnetism and thermalization
  in a spin-3 many-body dipolar lattice system}},}\ }\href@noop {} {\bibfield
  {journal} {\bibinfo  {journal} {arXiv:1803.02628}\ } (\bibinfo {year}
  {2018})}\BibitemShut {NoStop}%
\bibitem [{\citenamefont {Tang}\ \emph {et~al.}(2018)\citenamefont {Tang},
  \citenamefont {Kao}, \citenamefont {Li}, \citenamefont {Seo}, \citenamefont
  {Mallayya}, \citenamefont {Rigol}, \citenamefont {Gopalakrishnan},\ and\
  \citenamefont {Lev}}]{Tang:2017uu}%
  \BibitemOpen
  \bibfield  {author} {\bibinfo {author} {\bibfnamefont {Y.}~\bibnamefont
  {Tang}}, \bibinfo {author} {\bibfnamefont {W.}~\bibnamefont {Kao}}, \bibinfo
  {author} {\bibfnamefont {K.}~\bibnamefont {Li}}, \bibinfo {author}
  {\bibfnamefont {S.}~\bibnamefont {Seo}}, \bibinfo {author} {\bibfnamefont
  {K.}~\bibnamefont {Mallayya}}, \bibinfo {author} {\bibfnamefont
  {M.}~\bibnamefont {Rigol}}, \bibinfo {author} {\bibfnamefont
  {S.}~\bibnamefont {Gopalakrishnan}}, \ and\ \bibinfo {author} {\bibfnamefont
  {B.~L.}\ \bibnamefont {Lev}},\ }\bibfield  {title} {\enquote {\bibinfo
  {title} {{Thermalization near integrability in a dipolar quantum Newton's
  cradle}},}\ }\href@noop {} {\  (\bibinfo {year} {2018})},\ \Eprint
  {http://arxiv.org/abs/arXiv:1707.07031; to appear in PRX} {arXiv:1707.07031;
  to appear in PRX} \BibitemShut {NoStop}%
\bibitem [{\citenamefont {Pedri}\ and\ \citenamefont
  {Santos}(2005)}]{Pedri:2005cc}%
  \BibitemOpen
  \bibfield  {author} {\bibinfo {author} {\bibfnamefont {P.}~\bibnamefont
  {Pedri}}\ and\ \bibinfo {author} {\bibfnamefont {L.}~\bibnamefont {Santos}},\
  }\bibfield  {title} {\enquote {\bibinfo {title} {Two-dimensional bright
  solitons in dipolar {Bose-Einstein} condensates},}\ }\href@noop {} {\bibfield
   {journal} {\bibinfo  {journal} {Phys. Rev. Lett.}\ }\textbf {\bibinfo
  {volume} {95}},\ \bibinfo {pages} {200404} (\bibinfo {year}
  {2005})}\BibitemShut {NoStop}%
\bibitem [{\citenamefont {Nath}\ \emph {et~al.}(2008)\citenamefont {Nath},
  \citenamefont {Pedri},\ and\ \citenamefont {Santos}}]{Nath:2008gt}%
  \BibitemOpen
  \bibfield  {author} {\bibinfo {author} {\bibfnamefont {R.}~\bibnamefont
  {Nath}}, \bibinfo {author} {\bibfnamefont {P.}~\bibnamefont {Pedri}}, \ and\
  \bibinfo {author} {\bibfnamefont {L.}~\bibnamefont {Santos}},\ }\bibfield
  {title} {\enquote {\bibinfo {title} {Stability of dark solitons in three
  dimensional dipolar {Bose-Einstein} condensates},}\ }\href@noop {} {\bibfield
   {journal} {\bibinfo  {journal} {Phys. Rev. Lett.}\ }\textbf {\bibinfo
  {volume} {101}},\ \bibinfo {pages} {210402} (\bibinfo {year}
  {2008})}\BibitemShut {NoStop}%
\bibitem [{\citenamefont {Lahaye}\ \emph {et~al.}(2009)\citenamefont {Lahaye},
  \citenamefont {Menotti}, \citenamefont {Santos}, \citenamefont {Lewenstein},\
  and\ \citenamefont {Pfau}}]{DipolarBosonsReview}%
  \BibitemOpen
  \bibfield  {author} {\bibinfo {author} {\bibfnamefont {T.}~\bibnamefont
  {Lahaye}}, \bibinfo {author} {\bibfnamefont {C.}~\bibnamefont {Menotti}},
  \bibinfo {author} {\bibfnamefont {L.}~\bibnamefont {Santos}}, \bibinfo
  {author} {\bibfnamefont {M.}~\bibnamefont {Lewenstein}}, \ and\ \bibinfo
  {author} {\bibfnamefont {T.}~\bibnamefont {Pfau}},\ }\bibfield  {title}
  {\enquote {\bibinfo {title} {The physics of dipolar bosonic quantum gases},}\
  }\href@noop {} {\bibfield  {journal} {\bibinfo  {journal} {Rep. Prog. Phys.}\
  }\textbf {\bibinfo {volume} {72}},\ \bibinfo {pages} {126401} (\bibinfo
  {year} {2009})}\BibitemShut {NoStop}%
\bibitem [{\citenamefont {Mulkerin}\ \emph {et~al.}(2013)\citenamefont
  {Mulkerin}, \citenamefont {van Bijnen}, \citenamefont {O'Dell}, \citenamefont
  {Martin},\ and\ \citenamefont {Parker}}]{Mulkerin:2013cs}%
  \BibitemOpen
  \bibfield  {author} {\bibinfo {author} {\bibfnamefont {B.~C.}\ \bibnamefont
  {Mulkerin}}, \bibinfo {author} {\bibfnamefont {R.~M.~W.}\ \bibnamefont {van
  Bijnen}}, \bibinfo {author} {\bibfnamefont {D.~H.~J.}\ \bibnamefont
  {O'Dell}}, \bibinfo {author} {\bibfnamefont {A.~M.}\ \bibnamefont {Martin}},
  \ and\ \bibinfo {author} {\bibfnamefont {N.~G.}\ \bibnamefont {Parker}},\
  }\bibfield  {title} {\enquote {\bibinfo {title} {Anisotropic and long-range
  vortex interactions in two-dimensional dipolar bose gases},}\ }\href@noop {}
  {\bibfield  {journal} {\bibinfo  {journal} {Phys. Rev. Lett.}\ }\textbf
  {\bibinfo {volume} {111}},\ \bibinfo {pages} {170402} (\bibinfo {year}
  {2013})}\BibitemShut {NoStop}%
\bibitem [{\citenamefont {Klawunn}\ \emph {et~al.}(2008)\citenamefont
  {Klawunn}, \citenamefont {Nath}, \citenamefont {Pedri},\ and\ \citenamefont
  {Santos}}]{Klawunn:2008ep}%
  \BibitemOpen
  \bibfield  {author} {\bibinfo {author} {\bibfnamefont {M.}~\bibnamefont
  {Klawunn}}, \bibinfo {author} {\bibfnamefont {R.}~\bibnamefont {Nath}},
  \bibinfo {author} {\bibfnamefont {P.}~\bibnamefont {Pedri}}, \ and\ \bibinfo
  {author} {\bibfnamefont {L.}~\bibnamefont {Santos}},\ }\bibfield  {title}
  {\enquote {\bibinfo {title} {Transverse instability of straight vortex lines
  in dipolar {Bose-Einstein} condensates},}\ }\href@noop {} {\bibfield
  {journal} {\bibinfo  {journal} {Phys. Rev. Lett.}\ }\textbf {\bibinfo
  {volume} {100}},\ \bibinfo {pages} {240403} (\bibinfo {year}
  {2008})}\BibitemShut {NoStop}%
\bibitem [{\citenamefont {Cooper}\ \emph {et~al.}(2005)\citenamefont {Cooper},
  \citenamefont {Rezayi},\ and\ \citenamefont {Simon}}]{Cooper:2005hn}%
  \BibitemOpen
  \bibfield  {author} {\bibinfo {author} {\bibfnamefont {N.~R.}\ \bibnamefont
  {Cooper}}, \bibinfo {author} {\bibfnamefont {E.~H.}\ \bibnamefont {Rezayi}},
  \ and\ \bibinfo {author} {\bibfnamefont {S.~H.}\ \bibnamefont {Simon}},\
  }\bibfield  {title} {\enquote {\bibinfo {title} {Vortex lattices in rotating
  atomic bose gases with dipolar interactions},}\ }\href@noop {} {\bibfield
  {journal} {\bibinfo  {journal} {Phys. Rev. Lett.}\ }\textbf {\bibinfo
  {volume} {95}},\ \bibinfo {pages} {200402} (\bibinfo {year}
  {2005})}\BibitemShut {NoStop}%
\bibitem [{\citenamefont {Cherng}\ and\ \citenamefont
  {Demler}(2009)}]{Cherng:2009bk}%
  \BibitemOpen
  \bibfield  {author} {\bibinfo {author} {\bibfnamefont {R.~W.}\ \bibnamefont
  {Cherng}}\ and\ \bibinfo {author} {\bibfnamefont {E.}~\bibnamefont
  {Demler}},\ }\bibfield  {title} {\enquote {\bibinfo {title} {Magnetoroton
  softening in rb spinor condensates with dipolar interactions},}\ }\href@noop
  {} {\bibfield  {journal} {\bibinfo  {journal} {Phys. Rev. Lett.}\ }\textbf
  {\bibinfo {volume} {103}},\ \bibinfo {pages} {185301} (\bibinfo {year}
  {2009})}\BibitemShut {NoStop}%
\bibitem [{\citenamefont {He}\ \emph {et~al.}(2008)\citenamefont {He},
  \citenamefont {Zhang}, \citenamefont {Zhang},\ and\ \citenamefont
  {Yi}}]{He:2008jw}%
  \BibitemOpen
  \bibfield  {author} {\bibinfo {author} {\bibfnamefont {L.}~\bibnamefont
  {He}}, \bibinfo {author} {\bibfnamefont {J.~N.}\ \bibnamefont {Zhang}},
  \bibinfo {author} {\bibfnamefont {Y.}~\bibnamefont {Zhang}}, \ and\ \bibinfo
  {author} {\bibfnamefont {S.}~\bibnamefont {Yi}},\ }\bibfield  {title}
  {\enquote {\bibinfo {title} {Stability and free expansion of a dipolar fermi
  gas},}\ }\href@noop {} {\bibfield  {journal} {\bibinfo  {journal} {Phys. Rev.
  A}\ }\textbf {\bibinfo {volume} {77}},\ \bibinfo {pages} {031605} (\bibinfo
  {year} {2008})}\BibitemShut {NoStop}%
\bibitem [{\citenamefont {Fregoso}\ \emph {et~al.}(2009)\citenamefont
  {Fregoso}, \citenamefont {Sun}, \citenamefont {Fradkin},\ and\ \citenamefont
  {Lev}}]{Fregoso:2009cc}%
  \BibitemOpen
  \bibfield  {author} {\bibinfo {author} {\bibfnamefont {B.~M.}\ \bibnamefont
  {Fregoso}}, \bibinfo {author} {\bibfnamefont {K.}~\bibnamefont {Sun}},
  \bibinfo {author} {\bibfnamefont {E.}~\bibnamefont {Fradkin}}, \ and\
  \bibinfo {author} {\bibfnamefont {B.~L.}\ \bibnamefont {Lev}},\ }\bibfield
  {title} {\enquote {\bibinfo {title} {Biaxial nematic phases in ultracold
  dipolar fermi gases},}\ }\href@noop {} {\bibfield  {journal} {\bibinfo
  {journal} {New J. Phys.}\ }\textbf {\bibinfo {volume} {11}},\ \bibinfo
  {pages} {103003--8} (\bibinfo {year} {2009})}\BibitemShut {NoStop}%
\bibitem [{\citenamefont {Zhang}\ and\ \citenamefont
  {Yi}(2009)}]{Zhang:2009bw}%
  \BibitemOpen
  \bibfield  {author} {\bibinfo {author} {\bibfnamefont {J.~N.}\ \bibnamefont
  {Zhang}}\ and\ \bibinfo {author} {\bibfnamefont {S.}~\bibnamefont {Yi}},\
  }\bibfield  {title} {\enquote {\bibinfo {title} {Fermi surface of a trapped
  dipolar fermi gas},}\ }\href@noop {} {\bibfield  {journal} {\bibinfo
  {journal} {Phys. Rev. A}\ }\textbf {\bibinfo {volume} {80}},\ \bibinfo
  {pages} {053614} (\bibinfo {year} {2009})}\BibitemShut {NoStop}%
\bibitem [{\citenamefont {Aikawa}\ \emph
  {et~al.}(2014{\natexlab{b}})\citenamefont {Aikawa}, \citenamefont {Baier},
  \citenamefont {Frisch}, \citenamefont {Mark}, \citenamefont {Ravensbergen},\
  and\ \citenamefont {Ferlaino}}]{Aikawa:2014gd}%
  \BibitemOpen
  \bibfield  {author} {\bibinfo {author} {\bibfnamefont {K.}~\bibnamefont
  {Aikawa}}, \bibinfo {author} {\bibfnamefont {S.}~\bibnamefont {Baier}},
  \bibinfo {author} {\bibfnamefont {A.}~\bibnamefont {Frisch}}, \bibinfo
  {author} {\bibfnamefont {M.}~\bibnamefont {Mark}}, \bibinfo {author}
  {\bibfnamefont {C.}~\bibnamefont {Ravensbergen}}, \ and\ \bibinfo {author}
  {\bibfnamefont {F.}~\bibnamefont {Ferlaino}},\ }\bibfield  {title} {\enquote
  {\bibinfo {title} {Observation of fermi surface deformation in a dipolar
  quantum gas},}\ }\href@noop {} {\bibfield  {journal} {\bibinfo  {journal}
  {Science}\ }\textbf {\bibinfo {volume} {345}},\ \bibinfo {pages} {1484}
  (\bibinfo {year} {2014}{\natexlab{b}})}\BibitemShut {NoStop}%
\bibitem [{\citenamefont {Sinha}\ and\ \citenamefont
  {Santos}(2007)}]{Sinha:2007gx}%
  \BibitemOpen
  \bibfield  {author} {\bibinfo {author} {\bibfnamefont {S.}~\bibnamefont
  {Sinha}}\ and\ \bibinfo {author} {\bibfnamefont {L.}~\bibnamefont {Santos}},\
  }\bibfield  {title} {\enquote {\bibinfo {title} {Cold dipolar gases in
  quasi-one-dimensional geometries},}\ }\href@noop {} {\bibfield  {journal}
  {\bibinfo  {journal} {Phys. Rev. Lett.}\ }\textbf {\bibinfo {volume} {99}},\
  \bibinfo {pages} {140406} (\bibinfo {year} {2007})}\BibitemShut {NoStop}%
\bibitem [{\citenamefont {Citro}\ \emph {et~al.}(2007)\citenamefont {Citro},
  \citenamefont {Orignac}, \citenamefont {De~Palo},\ and\ \citenamefont
  {Chiofalo}}]{Citro:2007gs}%
  \BibitemOpen
  \bibfield  {author} {\bibinfo {author} {\bibfnamefont {R.}~\bibnamefont
  {Citro}}, \bibinfo {author} {\bibfnamefont {E.}~\bibnamefont {Orignac}},
  \bibinfo {author} {\bibfnamefont {S.}~\bibnamefont {De~Palo}}, \ and\
  \bibinfo {author} {\bibfnamefont {M.~L.}\ \bibnamefont {Chiofalo}},\
  }\bibfield  {title} {\enquote {\bibinfo {title} {{Evidence of
  Luttinger-liquid behavior in one-dimensional dipolar quantum gases}},}\
  }\href@noop {} {\bibfield  {journal} {\bibinfo  {journal} {Phys. Rev. A}\
  }\textbf {\bibinfo {volume} {75}},\ \bibinfo {pages} {051602} (\bibinfo
  {year} {2007})}\BibitemShut {NoStop}%
\bibitem [{\citenamefont {Yi}\ \emph {et~al.}(2007)\citenamefont {Yi},
  \citenamefont {Li},\ and\ \citenamefont {Sun}}]{Yi:2007dt}%
  \BibitemOpen
  \bibfield  {author} {\bibinfo {author} {\bibfnamefont {S.}~\bibnamefont
  {Yi}}, \bibinfo {author} {\bibfnamefont {T.}~\bibnamefont {Li}}, \ and\
  \bibinfo {author} {\bibfnamefont {C.~P.}\ \bibnamefont {Sun}},\ }\bibfield
  {title} {\enquote {\bibinfo {title} {Novel quantum phases of dipolar bose
  gases in optical lattices},}\ }\href@noop {} {\bibfield  {journal} {\bibinfo
  {journal} {Phys. Rev. Lett.}\ }\textbf {\bibinfo {volume} {98}},\ \bibinfo
  {pages} {260405} (\bibinfo {year} {2007})}\BibitemShut {NoStop}%
\bibitem [{\citenamefont {Hafezi}\ \emph {et~al.}(2007)\citenamefont {Hafezi},
  \citenamefont {S{\o}rensen}, \citenamefont {Demler},\ and\ \citenamefont
  {Lukin}}]{Hafezi:2007gz}%
  \BibitemOpen
  \bibfield  {author} {\bibinfo {author} {\bibfnamefont {M.}~\bibnamefont
  {Hafezi}}, \bibinfo {author} {\bibfnamefont {A.~S.}\ \bibnamefont
  {S{\o}rensen}}, \bibinfo {author} {\bibfnamefont {E.}~\bibnamefont {Demler}},
  \ and\ \bibinfo {author} {\bibfnamefont {M.~D.}\ \bibnamefont {Lukin}},\
  }\bibfield  {title} {\enquote {\bibinfo {title} {{Fractional quantum Hall
  effect in optical lattices}},}\ }\href@noop {} {\bibfield  {journal}
  {\bibinfo  {journal} {Phys. Rev. A}\ }\textbf {\bibinfo {volume} {76}},\
  \bibinfo {pages} {023613} (\bibinfo {year} {2007})}\BibitemShut {NoStop}%
\bibitem [{\citenamefont {Liu}\ \emph {et~al.}(2015)\citenamefont {Liu},
  \citenamefont {Li}, \citenamefont {Yin},\ and\ \citenamefont
  {Liu}}]{Liu:2015kq}%
  \BibitemOpen
  \bibfield  {author} {\bibinfo {author} {\bibfnamefont {B.}~\bibnamefont
  {Liu}}, \bibinfo {author} {\bibfnamefont {X.}~\bibnamefont {Li}}, \bibinfo
  {author} {\bibfnamefont {L.}~\bibnamefont {Yin}}, \ and\ \bibinfo {author}
  {\bibfnamefont {W.~V.}\ \bibnamefont {Liu}},\ }\bibfield  {title} {\enquote
  {\bibinfo {title} {{Weyl Superfluidity in a three-dimensional dipolar fermi
  gas}},}\ }\href@noop {} {\bibfield  {journal} {\bibinfo  {journal} {Phys.
  Rev. Lett.}\ }\textbf {\bibinfo {volume} {114}},\ \bibinfo {pages} {045302}
  (\bibinfo {year} {2015})}\BibitemShut {NoStop}%
\bibitem [{\citenamefont {Fattori}\ \emph {et~al.}(2008)\citenamefont
  {Fattori}, \citenamefont {Roati}, \citenamefont {Deissler}, \citenamefont
  {D'Errico}, \citenamefont {Zaccanti}, \citenamefont {Jona-Lasinio},
  \citenamefont {Santos}, \citenamefont {Inguscio},\ and\ \citenamefont
  {Modugno}}]{Fattori:2008jj}%
  \BibitemOpen
  \bibfield  {author} {\bibinfo {author} {\bibfnamefont {M.}~\bibnamefont
  {Fattori}}, \bibinfo {author} {\bibfnamefont {G.}~\bibnamefont {Roati}},
  \bibinfo {author} {\bibfnamefont {B.}~\bibnamefont {Deissler}}, \bibinfo
  {author} {\bibfnamefont {C.}~\bibnamefont {D'Errico}}, \bibinfo {author}
  {\bibfnamefont {M.}~\bibnamefont {Zaccanti}}, \bibinfo {author}
  {\bibfnamefont {M.}~\bibnamefont {Jona-Lasinio}}, \bibinfo {author}
  {\bibfnamefont {L.}~\bibnamefont {Santos}}, \bibinfo {author} {\bibfnamefont
  {M.}~\bibnamefont {Inguscio}}, \ and\ \bibinfo {author} {\bibfnamefont
  {G.}~\bibnamefont {Modugno}},\ }\bibfield  {title} {\enquote {\bibinfo
  {title} {Magnetic dipolar interaction in a {Bose-Einstein} condensate atomic
  interferometer},}\ }\href@noop {} {\bibfield  {journal} {\bibinfo  {journal}
  {Phys. Rev. Lett.}\ }\textbf {\bibinfo {volume} {101}},\ \bibinfo {pages}
  {190405} (\bibinfo {year} {2008})}\BibitemShut {NoStop}%
\bibitem [{\citenamefont {Baym}\ and\ \citenamefont {Hatsuda}()}]{Baym09}%
  \BibitemOpen
  \bibfield  {author} {\bibinfo {author} {\bibfnamefont {G.}~\bibnamefont
  {Baym}}\ and\ \bibinfo {author} {\bibfnamefont {T.}~\bibnamefont {Hatsuda}},\
  }\href@noop {} {}\bibinfo {note} {Private communication.}\BibitemShut {Stop}%
\bibitem [{\citenamefont {Giovanazzi}\ \emph {et~al.}(2002)\citenamefont
  {Giovanazzi}, \citenamefont {G\"orlitz},\ and\ \citenamefont
  {Pfau}}]{Giovanazzi:2002}%
  \BibitemOpen
  \bibfield  {author} {\bibinfo {author} {\bibfnamefont {S.}~\bibnamefont
  {Giovanazzi}}, \bibinfo {author} {\bibfnamefont {A.}~\bibnamefont
  {G\"orlitz}}, \ and\ \bibinfo {author} {\bibfnamefont {T.}~\bibnamefont
  {Pfau}},\ }\bibfield  {title} {\enquote {\bibinfo {title} {Tuning the dipolar
  interaction in quantum gases},}\ }\href {\doibase
  10.1103/PhysRevLett.89.130401} {\bibfield  {journal} {\bibinfo  {journal}
  {Phys. Rev. Lett.}\ }\textbf {\bibinfo {volume} {89}},\ \bibinfo {pages}
  {130401} (\bibinfo {year} {2002})}\BibitemShut {NoStop}%
\bibitem [{\citenamefont {Chin}\ \emph {et~al.}(2010)\citenamefont {Chin},
  \citenamefont {Grimm}, \citenamefont {Julienne},\ and\ \citenamefont
  {Tiesinga}}]{Chin10_RMP}%
  \BibitemOpen
  \bibfield  {author} {\bibinfo {author} {\bibfnamefont {C.}~\bibnamefont
  {Chin}}, \bibinfo {author} {\bibfnamefont {R.}~\bibnamefont {Grimm}},
  \bibinfo {author} {\bibfnamefont {P.}~\bibnamefont {Julienne}}, \ and\
  \bibinfo {author} {\bibfnamefont {E.}~\bibnamefont {Tiesinga}},\ }\bibfield
  {title} {\enquote {\bibinfo {title} {Feshbach resonances in ultracold
  gases},}\ }\href@noop {} {\bibfield  {journal} {\bibinfo  {journal} {Rev.
  Mod. Phys.}\ }\textbf {\bibinfo {volume} {82}},\ \bibinfo {pages} {1225}
  (\bibinfo {year} {2010})}\BibitemShut {NoStop}%
\bibitem [{Note1()}]{Note1}%
  \BibitemOpen
  \bibinfo {note} {Control of $\epsilon $ via a Feshbach resonance is often
  difficult in many highly magnetic atomic isotopes due to the narrowness of
  the resonances~\cite {Kotochigova:2014kt}.}\BibitemShut {Stop}%
\bibitem [{\citenamefont {Hensler}\ \emph {et~al.}(2003)\citenamefont
  {Hensler}, \citenamefont {Werner}, \citenamefont {Griesmaier}, \citenamefont
  {Schmidt}, \citenamefont {G{\"o}rlitz}, \citenamefont {Pfau}, \citenamefont
  {Giovanazzi},\ and\ \citenamefont {Rza{\.{z}}ewski}}]{Hensler:2003}%
  \BibitemOpen
  \bibfield  {author} {\bibinfo {author} {\bibfnamefont {S.}~\bibnamefont
  {Hensler}}, \bibinfo {author} {\bibfnamefont {J.}~\bibnamefont {Werner}},
  \bibinfo {author} {\bibfnamefont {A.}~\bibnamefont {Griesmaier}}, \bibinfo
  {author} {\bibfnamefont {P.~O.}\ \bibnamefont {Schmidt}}, \bibinfo {author}
  {\bibfnamefont {A.}~\bibnamefont {G{\"o}rlitz}}, \bibinfo {author}
  {\bibfnamefont {T.}~\bibnamefont {Pfau}}, \bibinfo {author} {\bibfnamefont
  {S.}~\bibnamefont {Giovanazzi}}, \ and\ \bibinfo {author} {\bibfnamefont
  {K.}~\bibnamefont {Rza{\.{z}}ewski}},\ }\bibfield  {title} {\enquote
  {\bibinfo {title} {Dipolar relaxation in an ultra-cold gas of magnetically
  trapped chromium atoms},}\ }\href {\doibase 10.1007/s00340-003-1334-0}
  {\bibfield  {journal} {\bibinfo  {journal} {Applied Physics B}\ }\textbf
  {\bibinfo {volume} {77}},\ \bibinfo {pages} {765} (\bibinfo {year}
  {2003})}\BibitemShut {NoStop}%
\bibitem [{\citenamefont {Pasquiou}\ \emph {et~al.}(2010)\citenamefont
  {Pasquiou}, \citenamefont {Bismut}, \citenamefont {Beaufils}, \citenamefont
  {Crubellier}, \citenamefont {Mar\'echal}, \citenamefont {Pedri},
  \citenamefont {Vernac}, \citenamefont {Gorceix},\ and\ \citenamefont
  {Laburthe-Tolra}}]{Pasquiou:2010}%
  \BibitemOpen
  \bibfield  {author} {\bibinfo {author} {\bibfnamefont {B.}~\bibnamefont
  {Pasquiou}}, \bibinfo {author} {\bibfnamefont {G.}~\bibnamefont {Bismut}},
  \bibinfo {author} {\bibfnamefont {Q.}~\bibnamefont {Beaufils}}, \bibinfo
  {author} {\bibfnamefont {A.}~\bibnamefont {Crubellier}}, \bibinfo {author}
  {\bibfnamefont {E.}~\bibnamefont {Mar\'echal}}, \bibinfo {author}
  {\bibfnamefont {P.}~\bibnamefont {Pedri}}, \bibinfo {author} {\bibfnamefont
  {L.}~\bibnamefont {Vernac}}, \bibinfo {author} {\bibfnamefont
  {O.}~\bibnamefont {Gorceix}}, \ and\ \bibinfo {author} {\bibfnamefont
  {B.}~\bibnamefont {Laburthe-Tolra}},\ }\bibfield  {title} {\enquote {\bibinfo
  {title} {Control of dipolar relaxation in external fields},}\ }\href
  {\doibase 10.1103/PhysRevA.81.042716} {\bibfield  {journal} {\bibinfo
  {journal} {Phys. Rev. A}\ }\textbf {\bibinfo {volume} {81}},\ \bibinfo
  {pages} {042716} (\bibinfo {year} {2010})}\BibitemShut {NoStop}%
\bibitem [{\citenamefont {Burdick}\ \emph {et~al.}(2015)\citenamefont
  {Burdick}, \citenamefont {Baumann}, \citenamefont {Tang}, \citenamefont
  {Lu},\ and\ \citenamefont {Lev}}]{Burdick:2015bx}%
  \BibitemOpen
  \bibfield  {author} {\bibinfo {author} {\bibfnamefont {N.~Q.}\ \bibnamefont
  {Burdick}}, \bibinfo {author} {\bibfnamefont {K.}~\bibnamefont {Baumann}},
  \bibinfo {author} {\bibfnamefont {Y.}~\bibnamefont {Tang}}, \bibinfo {author}
  {\bibfnamefont {M.}~\bibnamefont {Lu}}, \ and\ \bibinfo {author}
  {\bibfnamefont {B.~L.}\ \bibnamefont {Lev}},\ }\bibfield  {title} {\enquote
  {\bibinfo {title} {{Fermionic Suppression of Dipolar Relaxation}},}\ }\href
  {\doibase 10.1103/PhysRevLett.114.023201} {\bibfield  {journal} {\bibinfo
  {journal} {Phys. Rev. Lett.}\ }\textbf {\bibinfo {volume} {114}},\ \bibinfo
  {pages} {023201} (\bibinfo {year} {2015})}\BibitemShut {NoStop}%
\bibitem [{\citenamefont {Tang}\ \emph {et~al.}(2015)\citenamefont {Tang},
  \citenamefont {Burdick}, \citenamefont {Baumann},\ and\ \citenamefont
  {Lev}}]{Tang2015}%
  \BibitemOpen
  \bibfield  {author} {\bibinfo {author} {\bibfnamefont {Y.}~\bibnamefont
  {Tang}}, \bibinfo {author} {\bibfnamefont {N.~Q.}\ \bibnamefont {Burdick}},
  \bibinfo {author} {\bibfnamefont {K.}~\bibnamefont {Baumann}}, \ and\
  \bibinfo {author} {\bibfnamefont {B.~L.}\ \bibnamefont {Lev}},\ }\bibfield
  {title} {\enquote {\bibinfo {title} {{Bose--Einstein condensation of
  $^{162}$Dy and $^{160}$Dy}},}\ }\href@noop {} {\bibfield  {journal} {\bibinfo
   {journal} {New J. Phys.}\ }\textbf {\bibinfo {volume} {17}},\ \bibinfo
  {pages} {045006} (\bibinfo {year} {2015})}\BibitemShut {NoStop}%
\bibitem [{Note2()}]{Note2}%
  \BibitemOpen
  \bibinfo {note} {The error in this number is quoted as the worst case error.
  This worst case occurs at $\varphi =90^\circ $, since the uncertainty in
  $B_\protect \text {rot}$ is larger than in $B_z$.}\BibitemShut {Stop}%
\bibitem [{\citenamefont {Baumann}\ \emph {et~al.}(2014)\citenamefont
  {Baumann}, \citenamefont {Burdick}, \citenamefont {Lu},\ and\ \citenamefont
  {Lev}}]{Baumann:2014ey}%
  \BibitemOpen
  \bibfield  {author} {\bibinfo {author} {\bibfnamefont {K.}~\bibnamefont
  {Baumann}}, \bibinfo {author} {\bibfnamefont {N.~Q.}\ \bibnamefont
  {Burdick}}, \bibinfo {author} {\bibfnamefont {M.}~\bibnamefont {Lu}}, \ and\
  \bibinfo {author} {\bibfnamefont {B.~L.}\ \bibnamefont {Lev}},\ }\bibfield
  {title} {\enquote {\bibinfo {title} {{Observation of low-field Fano-Feshbach
  resonances in ultracold gases of dysprosium}},}\ }\href {\doibase
  10.1103/PhysRevA.89.020701} {\bibfield  {journal} {\bibinfo  {journal} {Phys.
  Rev. A}\ }\textbf {\bibinfo {volume} {89}},\ \bibinfo {pages} {020701}
  (\bibinfo {year} {2014})}\BibitemShut {NoStop}%
\bibitem [{\citenamefont {Martin}\ \emph {et~al.}(1978)\citenamefont {Martin},
  \citenamefont {Zalubas},\ and\ \citenamefont {Hagan}}]{Martin:1978}%
  \BibitemOpen
  \bibfield  {author} {\bibinfo {author} {\bibfnamefont {W.~C.}\ \bibnamefont
  {Martin}}, \bibinfo {author} {\bibfnamefont {R.}~\bibnamefont {Zalubas}}, \
  and\ \bibinfo {author} {\bibfnamefont {L.}~\bibnamefont {Hagan}},\
  }\href@noop {} {\emph {\bibinfo {title} {Atomic Energy Levels--The Rare Earth
  Elements}}}\ (\bibinfo  {publisher} {NSRDS-NBS, \textbf{60}},\ \bibinfo
  {address} {Washington, D.C.},\ \bibinfo {year} {1978})\BibitemShut {NoStop}%
\bibitem [{\citenamefont {Castin}\ and\ \citenamefont
  {Dum}(1996)}]{Castin:1996}%
  \BibitemOpen
  \bibfield  {author} {\bibinfo {author} {\bibfnamefont {Y.}~\bibnamefont
  {Castin}}\ and\ \bibinfo {author} {\bibfnamefont {R.}~\bibnamefont {Dum}},\
  }\bibfield  {title} {\enquote {\bibinfo {title} {{Bose-Einstein} condensates
  in time dependent traps},}\ }\href {\doibase 10.1103/PhysRevLett.77.5315}
  {\bibfield  {journal} {\bibinfo  {journal} {Phys. Rev. Lett.}\ }\textbf
  {\bibinfo {volume} {77}},\ \bibinfo {pages} {5315} (\bibinfo {year}
  {1996})}\BibitemShut {NoStop}%
\bibitem [{\citenamefont {Tang}\ \emph {et~al.}(2016)\citenamefont {Tang},
  \citenamefont {Sykes}, \citenamefont {Burdick}, \citenamefont {DiSciacca},
  \citenamefont {Petrov},\ and\ \citenamefont {Lev}}]{Tang:2016prl}%
  \BibitemOpen
  \bibfield  {author} {\bibinfo {author} {\bibfnamefont {Y.}~\bibnamefont
  {Tang}}, \bibinfo {author} {\bibfnamefont {A.~G.}\ \bibnamefont {Sykes}},
  \bibinfo {author} {\bibfnamefont {N.~Q.}\ \bibnamefont {Burdick}}, \bibinfo
  {author} {\bibfnamefont {J.~M.}\ \bibnamefont {DiSciacca}}, \bibinfo {author}
  {\bibfnamefont {D.~S.}\ \bibnamefont {Petrov}}, \ and\ \bibinfo {author}
  {\bibfnamefont {B.~L.}\ \bibnamefont {Lev}},\ }\bibfield  {title} {\enquote
  {\bibinfo {title} {Anisotropic expansion of a thermal dipolar bose gas},}\
  }\href {\doibase 10.1103/PhysRevLett.117.155301} {\bibfield  {journal}
  {\bibinfo  {journal} {Phys. Rev. Lett.}\ }\textbf {\bibinfo {volume} {117}},\
  \bibinfo {pages} {155301} (\bibinfo {year} {2016})}\BibitemShut {NoStop}%
\bibitem [{Note3()}]{Note3}%
  \BibitemOpen
  \bibinfo {note} {We were unable to measure lifetimes at rotation rates faster
  than $\sim $1~kHz due to power supply limitations and increasingly large eddy
  currents.}\BibitemShut {Stop}%
\bibitem [{\citenamefont {Kotochigova}(2014)}]{Kotochigova:2014kt}%
  \BibitemOpen
  \bibfield  {author} {\bibinfo {author} {\bibfnamefont {S.}~\bibnamefont
  {Kotochigova}},\ }\bibfield  {title} {\enquote {\bibinfo {title}
  {{Controlling interactions between highly magnetic atoms with Feshbach
  resonances}},}\ }\href@noop {} {\bibfield  {journal} {\bibinfo  {journal}
  {Rep. Prog. Phys.}\ }\textbf {\bibinfo {volume} {77}},\ \bibinfo {pages}
  {093901} (\bibinfo {year} {2014})}\BibitemShut {NoStop}%
\end{thebibliography}
%

\end{document}